\documentclass[aps,prl,showpacs,floatfix,preprint,superscriptaddress]{revtex4-1}

\pdfoutput=1

\usepackage{graphicx,color}
\usepackage{amssymb,amsthm,amsmath,hyperref}
\begin{document}
\title{Dominant role of processing temperature in electric field induced superconductivity in layered ZrNBr}

\author{Xinmin Wang}
\affiliation{Institute of Physics and Beijing National Laboratory for Condensed Matter Physics, Chinese Academy of Sciences, Beijing 100190, China} 
\affiliation{School of Physical Sciences, University of Chinese Academy of Sciences, Beijing 100049, China}
\author{Shuai Zhang}
\email{szhang@iphy.ac.cn}
\affiliation{Institute of Physics and Beijing National Laboratory for Condensed Matter Physics, Chinese Academy of Sciences, Beijing 100190, China} 
\author{Huanyan Fu}
\affiliation{Institute of Physics and Beijing National Laboratory for Condensed Matter Physics, Chinese Academy of Sciences, Beijing 100190, China} 
\affiliation{School of Physics and Electronics, Shandong Normal University, Jinan 250014, China}
\author{Moran Gao}
\affiliation{Institute of Physics and Beijing National Laboratory for Condensed Matter Physics, Chinese Academy of Sciences, Beijing 100190, China} 
\affiliation{School of Physical Sciences, University of Chinese Academy of Sciences, Beijing 100049, China}
\author{Zhian Ren}
\affiliation{Institute of Physics and Beijing National Laboratory for Condensed Matter Physics, Chinese Academy of Sciences, Beijing 100190, China}
\affiliation{School of Physical Sciences, University of Chinese Academy of Sciences, Beijing 100049, China}
\affiliation{Collaborative Innovation Center of Quantum Matter, Beijing 100190, China}
\author{Genfu Chen}
\affiliation{Institute of Physics and Beijing National Laboratory for Condensed Matter Physics, Chinese Academy of Sciences, Beijing 100190, China}
\affiliation{School of Physical Sciences, University of Chinese Academy of Sciences, Beijing 100049, China}
\affiliation{Collaborative Innovation Center of Quantum Matter, Beijing 100190, China}

\begin{abstract}
Recently, as a novel technique, electronic double-layer transistors (EDLTs) with ionic liquids have shown strong potential for tuning the electronic states of correlated systems. EDLT induced local carrier doping can always lead to dramatic changes in physical properties when compared to parent materials, e.g., insulating-superconducting transition. Generally, the modification of gate voltage ($V_{\rm G}$) in EDLT devices produces a direct change on the doping level, whereas the processing temperature ($T_{\rm G}$) at which $V_{\rm G}$ is applied only assists the tuning process. Here, we report that the processing temperature $T_{\rm G}$ plays a dominant role in the electric field induced superconductivity in layered ZrNBr. When applying $V_{\rm G}$ at $T_{\rm G}\geq$ 250 K, the induced superconducting (SC) state permanently remains in the material, which is confirmed in the zero resistance and diamagnetism after long-time relaxation at room temperature and/or by applying reverse voltage, whereas the solid/liquid interface induced reversible insulating-SC transition occurs at $T_{\rm G}\leq$ 235 K. These experimental facts support another electrochemical mechanism that electric field induced partial deintercalation of Br ions could cause permanent electron doping to the system. Our findings in this study will extend the potential of electric fields for tuning bulk electronic states in low-dimension systems.
\end{abstract}
\maketitle


As one of the most effective methods of controlling electric fields, EDLT devices with ionic liquid can induce exotic states in strongly correlated parent materials upon modifying local carrier density. The EDLT induced insulating-SC transition\cite{UenoNM2008,YeNM2009,YeScience2012,SaitoNC2018}, full phase diagram through the charge density wave (CDW) and superconductivity in 1T-TiSe$_2$\cite{LiNature2016}, and enhancement of SC transition temperature\cite{LeiPRL2016} have attracted remarkable attention in material science. In general, applying $V_{\rm G}$s causes polarization of the cations and anions in ionic liquid, which can lead to the accumulation of the corresponding carriers on the crystal surface. Thus, the induced SC states mainly form in a rather thin layer with a thickness of 1 - 2 nm as reported in previous studies\cite{UenoNM2008,YeNM2009}. The carrier density equilibrates again in the material after releasing the $V_{\rm G}$ at temperatures higher than the melting point of the selected liquid dielectric, leading to the disappearance of the induced SC phase. Another characteristic of the general EDLT device is that modifying $V_{\rm G}$ produces a direct change on local carrier doping levels, whereas controlling the temperature $T_{\rm G}$ causes significant influence not on the tuning process but on the electrochemical window (EW) of the selected ionic liquid. When applying the same $V_{\rm G}$ at different temperatures, the solid/liquid interface results in a certain doping level as reaching the equilibrium between $V_{\rm G}$ and polarized dielectric. A relatively higher temperature could improve the mobility of ions and decrease the response time for the equilibrium\cite{YuanAFM2009}. The leakage current between gate and source electrodes increases with increasing temperature.

In recent studies, electric field is capable of tuning the bulk property by controlling ions rather than the local carriers governed in EDLT devices. Electrolyte gating of epitaxial thin film of VO$_2$ suppresses the metal-insulator transition and stabilizes the metallic phase through the electric field-induced creation of oxygen vacancy\cite{JeongScience2013}. The transformation between the three phases SrCoO$_{3-\delta}$, SrCoO$_{2.5}$, and HSrCoO$_{2.5}$ can be induced through the control of the insertion and extraction of oxygen and hydrogen ions independently in an electric field\cite{LuNature2017}. On the other hand, the developed control of the electric field makes it possible to access a complete set of competing electronic phases from band insulators and superconductors, to a reentrant insulator on the monolayers of WS$_2$\cite{LuPNAS2018}. 

Here, we present a temperature $T_{\rm G}$ dominated evolution of the electric field induced superconductivity in layered ZrNBr with a hexagonal [Zr$_2$N$_2$] layer. The parent ZrNBr is an insulator with an energy gap of $\sim$3 eV, which turns into a superconductor with $T_c=$ 14 K upon intercalation\cite{ArroyoIJNM2000}. When applying proper $V_{\rm G}$s at $T_{\rm G}s\geq$ 250 K, the induced SC state remains permanently in the material as confirmed in zero resistance and diamagnetism after long-time relaxation without $V_{\rm G}$ at room temperature and/or applying reverse voltage, whereas a reversible insulating-SC transition occurs in the gating process with $T_{\rm G}\leq$ 235 K. These facts strongly imply that another electrochemical mechanism such as electric field induced partial deintercalation of Br ions is a valid explanation of the permanent superconductivity. A hidden SC transition at 11 K is found to coexist with the primary transition at 14 K upon systematically modifying of $T_{\rm G}$ and $V_{\rm G}$.

\begin{figure}
\centering
\includegraphics[width=1\linewidth]{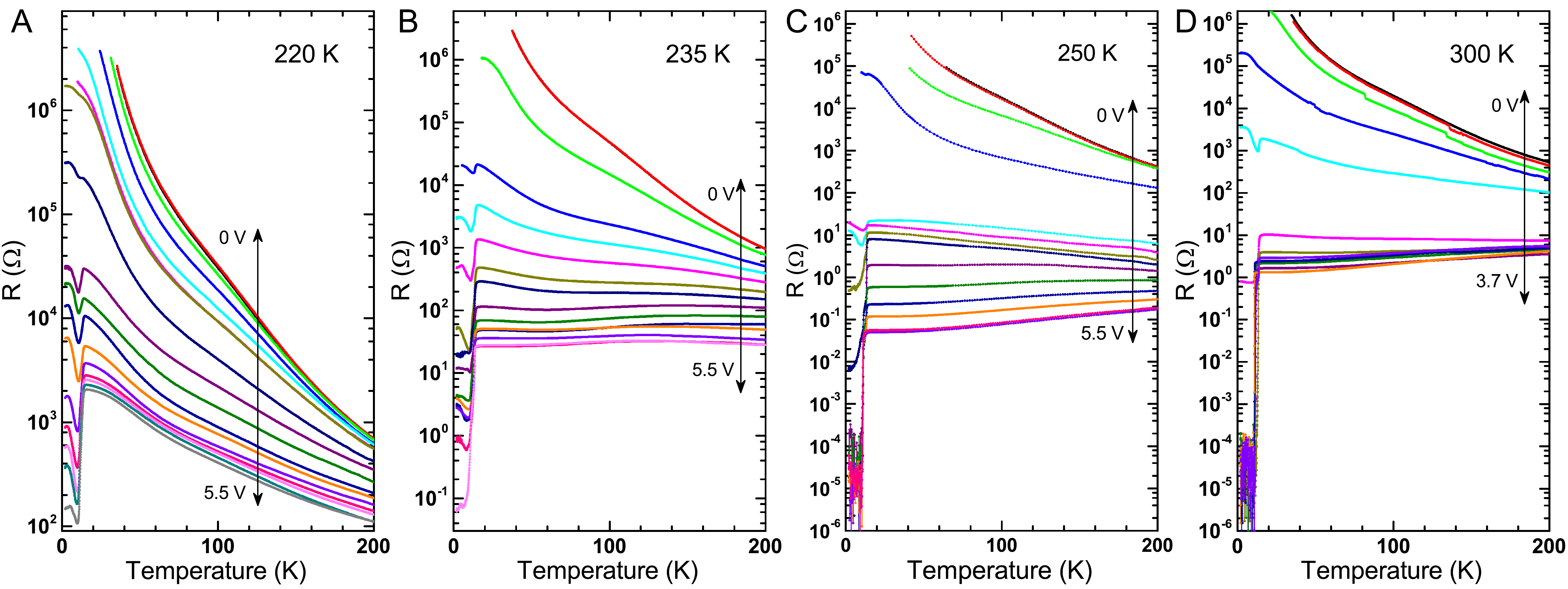}
\caption{Temperature dependence of resistance $R(T)$ induced by different $V_{\rm G}$s applied at 220 ($A$), 235 ($B$), 250 ($C$), and 300 K ($D$), respectively. For 220, 235, and 250 K, the $V_{\rm G}$s were applied from 0, 1.5, 2 to 5.5 V, whereas $V_{\rm G}$s up to 3.7 V were applied at 300 K. The step between 2 and 5.5 V was 0.2 - 0.3 V. For $T_{\rm G}=$ 300 K, $V_{\rm G}=$ 0, 1, 1.5, 2, 2.3, 2.5, 2.7, 3, 3.5, and 3.7 V were applied.}
\label{fig1}
\end{figure}

Figure \ref{fig1}$A$, $B$, $C$ and $D$ show the temperature dependence of the resistance $R(T)$ in different $V_{\rm G}$s applied at 220, 235, 250, and 300 K, respectively. For all temperatures in the range of 220 to 300 K, the absolute resistance at normal state decreases monotonically with increasing $V_{\rm G}$ (Supporting Information, section 1), implying that the carrier doping level is effectively controlled by $V_{\rm G}$. For a $T_{\rm G}=$ of 220 K, $R$ shows a transition around 14 K with magnitude of 10$^5$ $\Omega$ at 3 V. As $V_{\rm G}$ increases up to 5.5 V, low-temperature $R(T)$ decreases to a level of 10$^2$ $\Omega$ with the same transition temperature. The coexistence of a SC state and bulk insulating/semiconducting characteristics indicates that relatively low carrier doping leads to tiny amounts of induced SC components in the transport channel. $R(T)$ with different $V_{\rm G}$s applied at 235 K shows a similar overall tendency (Fig. \ref{fig1}$B$). A metallic feature, resistance decreasing with decreasing temperature, is present in low-temperature $R(T)$ ($<$ 150 K) when applying $V_{\rm G}$ up to 5 and 5.5 V, and the absolute $R(T)$ is of order $10^{-1} \Omega$ below $T_c$, although this value is still not considered as zero resistance. 

With increasing temperature $T_{\rm G}$ up to 250 and 300 K, the zero resistance ($10^{-5}\sim10^{-6}$ $\Omega$) is present at $V_{\rm G}\geq$ 2.5 V. However, the metallic $R(T)$ appears when applying $V_{\rm G}\geq$ 3.3 and 3.0 V at $T_{\rm G}=$ 250 and 300 K, respectively. A continuous change from insulating to metallic characteristics in bulk $R(T)$ can be seen clearly in the linear scale (Supporting Information, section 1). As we will later discuss, the overall effect of raising temperature $T_{\rm G}$ is shown not only in the appearance of zero resistance and metallic behavior in $R(T)$, but also in the induced permanent superconductivity. These facts strongly imply that the same $V_{\rm G}$ applied at different temperatures results in different electronic states. The processing temperature effect cannot be interpreted in the normal sense of EDLT.

\begin{figure}
\centering
\includegraphics[width=1\linewidth]{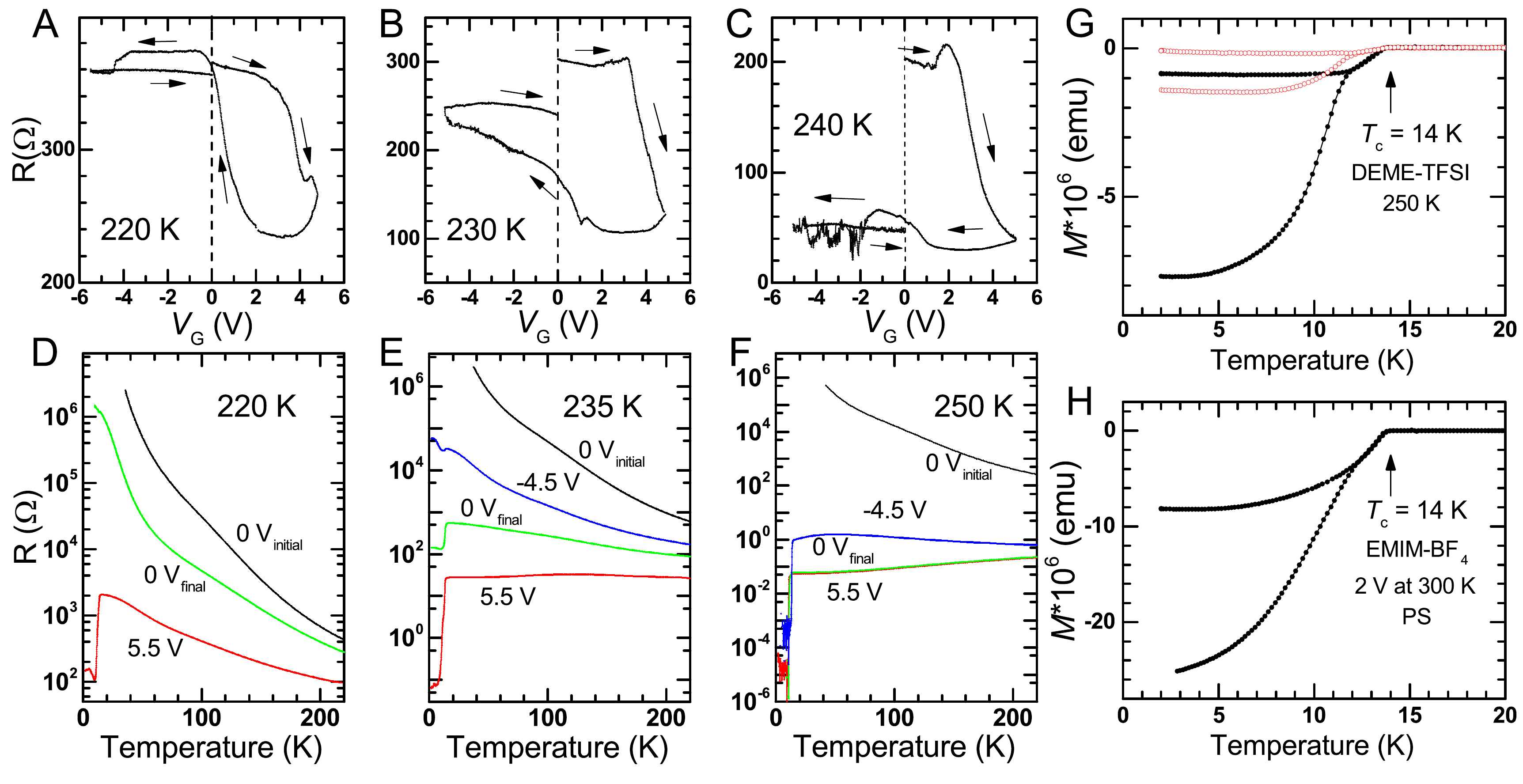}
\caption{Reversible and irreversible change between insulating and superconducting states. Isothermal $V_{\rm G}$ dependence of resistance measured at 220 K ($A$), 230 K ($B$), and 240 K ($C$). There are four quarters when sweeping $V_{\rm G}$ in one full loop; 0 V $\rightarrow$ $+5$ V $\rightarrow$ 0 V $\rightarrow$ $-5$ V $\rightarrow$ 0 V. The $R(T)$ curves measured after applying several critical $V_{\rm G}$s (0 V$_{\rm initial}$, +5 V, 0 V$_{\rm final}$, and $-4.5$ V) at 220 K ($D$), 235 K ($E$), and 250 K ($F$) show reversible, intermediate, and irreversible change between insulating and superconducting states. Diamagnetism was confirmed at the temperature corresponding to the electrical field induced SC transition ($G$ and $H$). Temperature dependence of $M(T)$ with $H=$ 10 Oe was measured on the same single crystals after completing electrical transport measurements at $T_{\rm G}=$ 250 K (\#1, solid circles in $G$), which is compared with the $M(T)$ after annealing at 120 degree for 3 days (open circles in $G$). $M(T)$ with $T_c=$ 14 K was obtained on a polycrystalline ZrNBr as applying $V_{\rm G}=$ 2 V at room temperature ($H$), in which EMIM-BF$_4$ was selected as the gate dielectric.}
\label{fig4}
\end{figure}

To further understand the effect of applying $V_{\rm G}$ at different temperatures, we measured the isothermal resistance when scanning $V_{\rm G}$. $V_{\rm G}$ was swept with 5 mV/s at several selected temperatures. A full $V_{\rm G}$ scanning loop consists of four quarters (0 V $\rightarrow$ $+5$ V $\rightarrow$ 0 V $\rightarrow$ $-5$ V $\rightarrow$ 0 V). When increasing $V_{\rm G}$ from 0 to 5 V at 220 K, $R$ decreases monotonically as shown in Fig. \ref{fig4}$A$. After the former two quarters, $R$ is almost recovered to the initial value at 0 V. Applying negative $V_{\rm G}$ causes a weak hysteresis of $R$. The perfect recovery of $R$ in the former two quarters provides direct evidence that the induced high-conductance state disappears immediately upon releasing the $V_{\rm G}$, which is consistent with the general EDLT effect. When increasing the temperature $T_{\rm G}$ to 230 K (Fig. \ref{fig4}$B$), $R$ shows a decrease of near 44\% and 21\% after the former two quarters and a full loop, respectively. Furthermore, a dramatic decrease (76\%) has been confirmed when scanning $V_{\rm G}$ at 240 K (Fig. \ref{fig4}$C$). It appears that the completely irreversible change in resistance occurs at temperatures above $\sim$235 K, which may suggest that another temperature dominated mechanism besides normal EDLT effect is present in the system.

Correspondingly, the reversible and irreversible changes between insulating and SC states were investigated using $R(T)$ in several critical $V_{\rm G}$s applied at this temperature region as shown in Fig. \ref{fig4}$D$, $E$ and $F$. The process of applying these $V_{\rm G}$s is described in Methods. As $T_{\rm G}=$ 220K, $R(T)$ with final 0 V shows the insulating behavior, which is very close to the initial state. Such a perfect recovery from SC to insulating state is significantly in agreement with the isothermal $V_{\rm G}$ dependence of resistance (Fig. \ref{fig4}$A$). Compared to $+5.5$ V applied at 235 K (Fig. \ref{fig4}$E$), final 0 V could only restore $R(T)$ back to $10^{2}$ $\Omega$ with a steep transition at 14 K. Further application of $-4.5$ V could restore the $R(T)$ back to an insulator-like state. A rather weak and clear drop can still be confirmed around 14 K, implying that partial induced SC states remain in the material. As temperature $T_{\rm G}$ increases to 250 K (Fig. \ref{fig4}$F$), final 0 V causes no obvious change compared to $+5.5$ V. The application of $-4.5$ V leads to a weak change from metallic to a weak semiconductor-like state with a clear SC transition, implying that the electric field induced superconductivity remains permanently in the system. All these experimental facts suggest that the reversible and irreversible changes between insulating and SC states strongly depend on the temperature $T_{\rm G}$, supporting again that the permanent superconductivity is not induced by the general EDLT effect but some other mechanism such as electric field induced partial deintercalation of Br. Similar electric field induced vacancy formation was observed in VO$_2$\cite{NakanoNature2012,JeongScience2013}. Furthermore, it was reported that the VO$_2$ films were seriously damaged by the application of a rather low $V_{\rm G}$ (2 V) at room temperature for several hours\cite{ZhouJAP2012}, implying that the processing time may be another important factor as well as $T_{\rm G}$ and $V_{\rm G}$.

The long-time relaxation and/or application of reverse $V_{\rm G}$ already proves that the superconductivity induced at $T_{\rm G}\geq 250$ K is robust as observed in the intercalated superconductors. In order to confirm the permanent characteristic, the magnetic responses of the same single crystal were measured after completing the electrical transport measurements. After applying $V_{\rm G}$ at 220 and 235 K, no SC transition was observed in the temperature dependence of magnetization $M(T)$. As shown in Fig. \ref{fig4}$E$, a weak drop is confirmed at approximately 14 K with huge resistance, This suggests that only a small amount of the induced SC components remain in the system after applying $-4.5$ V. It is reasonable that such a weak SC signal can scarcely be detected in $M(T)$. As $T_{\rm G}$ increases up to higher than 250 K, the primary SC transition at 14 K is confirmed in $M(T)$ (Fig. \ref{fig4}$G$). Furthermore, one can see that the diamagnetism is clearly weakened by a three day annealing at 120 $^\circ$C, although $T_c$ shows no apparent shift (open circles in Fig. \ref{fig4}$G$). The apparent annealing effect suggests that the induced SC phase is sensitive to high-temperature treatment. Fig. \ref{fig4}$H$ shows the $M(T)$ measured on a polycrystalline sample using a different gate dielectric EMIM-BF$_4$. When applying $V_{\rm G}=$ 2 V at room temperature, a clear transition is also confirmed at 14 K, strongly implying that the electric field induced permanent SC state is independent of the gate dielectric. 

\begin{figure}
\centering
\includegraphics[width=1\linewidth]{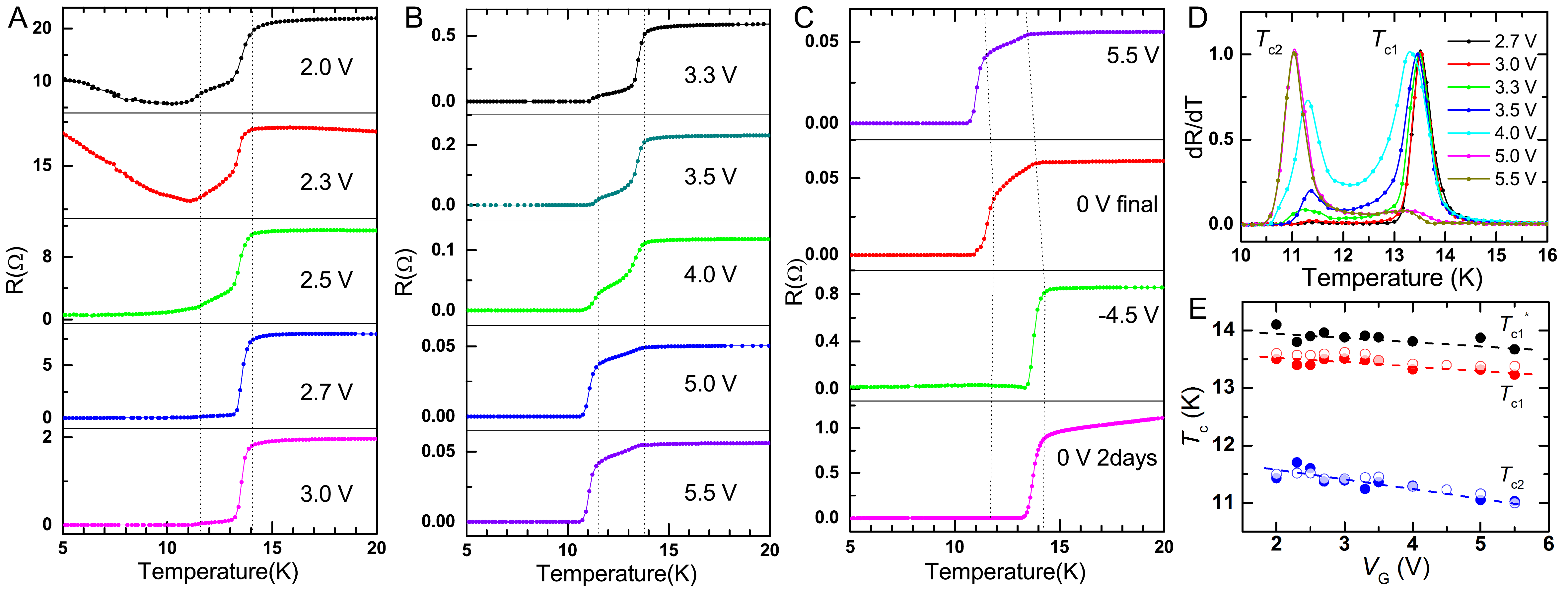}
\caption{$V_{\rm G}$ dependence of superconducting transition temperatures observed when applying $V_{\rm G}$ at 250 K. $R(T)$ in an expanded scale for $V_{\rm G}=$ 2.0-3.0 V ($A$), 3.3-5.5 V ($B$), and several critical $V_{\rm G}$ ($C$). Normalized $dR/dT$ is used to determine $T_{c1}$ and $T_{c2}$ ($D$). $V_{\rm G}$ dependence of $T_{c1}$ and $T_{c2}$ is compared with $T_{c}^*$ determined in a different method ($E$). The filled and open circles represent $T_c$ values obtained in two verification experiments on a single crystal $\#1$ and $\#2$.}
\label{fig2}
\end{figure}

On the other hand, we also explored the $V_{\rm G}$ dependence of the induced SC transition temperature. The $R(T)$ measured after applying different $V_{\rm G}$s at 250 K is selected as a typical example. When applying $V_{\rm G}$ from 2.0 to 2.7 V (Fig. \ref{fig2}$A$), a primary SC transition together with another weak drop was observed around 14 and 11 K, which are named $T_{c1}$ and $T_{c2}$, respectively. As mentioned previously, zero resistance was realized by increasing $V_{\rm G}$ to higher than 2.7 V. Thus, we reasonably consider that the induced SC states with $V_{\rm G}\leq 2.5$ V have rather poor homogeneity. By increasing $V_{\rm G}$ from 2.7 to 5.5 V (Fig. \ref{fig2}$B$), the primary transition at $T_{c1}$ is slightly weakened, and the transition around $T_{c2}$ becomes gradually obvious. Increasing $V_{\rm G}$ results in the decreasing of SC components with $T_{c1}$ as well as increasing the contribution from SC states with $T_{c2}$. Furthermore, long-time relaxation without voltage at 250 K appears to cause a weak reverse effect (Fig. \ref{fig2}$C$). Surprisingly, the transition at $T_{c2}$ disappears immediately when applying negative voltage ($-4.5$ V). A two-day relaxation at room temperature causes no apparent change on the electric field induced SC transition at $T_{c1}$ or the bulk metallic characteristic, except for a slightly increased resistance.

To quantitatively characterize the $V_{\rm G}$ dependence of the two induced SC transitions, we determine $T_{c1}$ and $T_{c2}$ as the temperatures corresponding to the peaks observed in the normalized $dR/dT$ (Fig. \ref{fig2}$D$). The two transitions can be clearly seen in $dR/dT$, which displays an alternating change between $T_{c1}$ and $T_{c2}$ as discussed previously. The SC states with $T_{c1}$ and $T_{c2}$ dominate at the low and high $V_{\rm G}$, respectively. As shown in Fig. \ref{fig2}$E$, the $T_{c1}$ and $T_{c2}$ values obtained at different $V_{\rm G}$s are compared with $T_{c1}^*$ determined using conventional extrapolation methods. Both $T_{c1}$ and $T_{c2}$ show a weak $V_{\rm G}$ dependence; $T_c$ decreases with increasing $V_{\rm G}$. The $V_{\rm G}$ dependence of $T_c$ is consistent with the $T_c$ dependence of doping levels observed in bulk intercalated superconductors such as Li$_xM_y$HfNCl\cite{TakanoPRL2008} and $AE_xM_y$HfNCl\cite{ZhangSUST20131}. The $V_{\rm G}$ dependence of $T_{c2}$ (open circles in Fig. \ref{fig2}$E$) suggests that the anomaly at $T_{c2}$ is not a random SC transition caused by non-homogeneity but a hidden SC phase. Such results have high reproducibility as confirmed in another single crystal (Supporting Information, section 3). However, the absence of $T_{c2}$ in $M(T)$ and $R(T)$ after applying negative $V_{\rm G}$ indicates the metastable characteristic of the SC states with $T_{c2}$.

\begin{figure}
\centering
\includegraphics[width=1\linewidth]{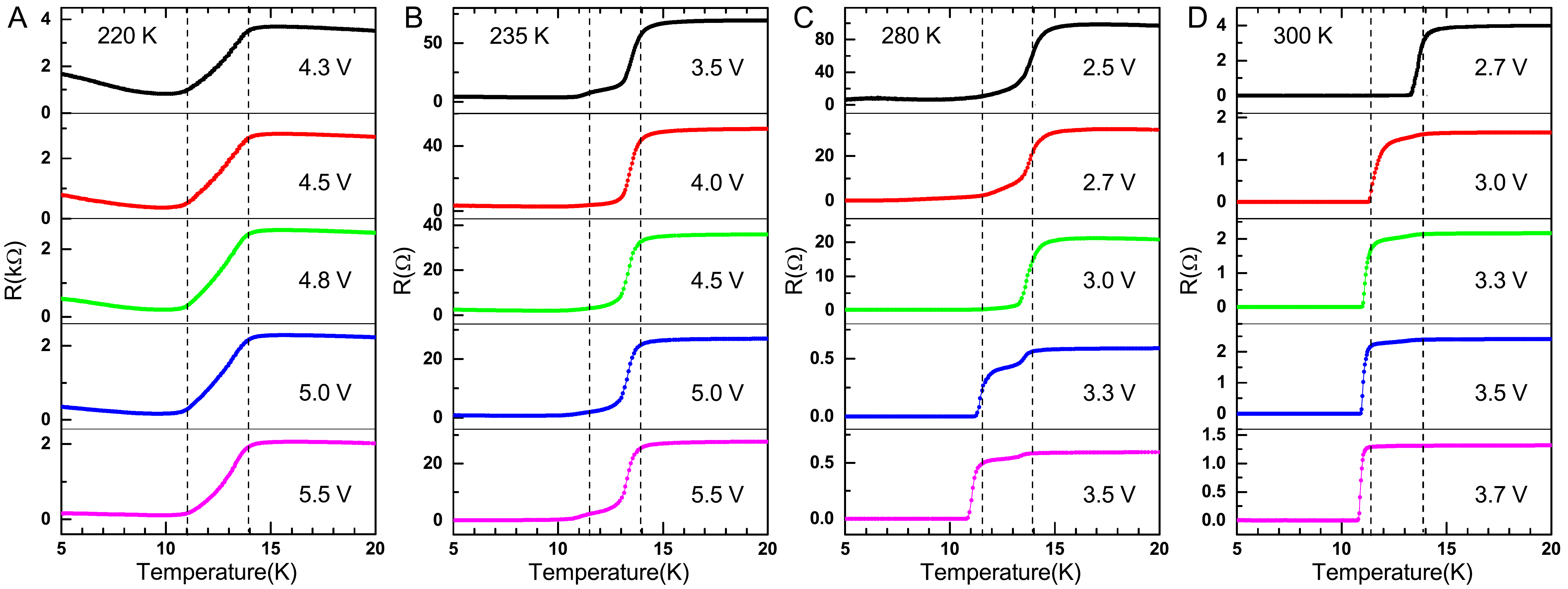}
\caption{$V_{\rm G}$ dependence of transition temperatures when applying $V_{\rm G}$s at 220 K ($A$), 235 K ($B$), 280 K ($C$), and 300 K($D$). The dashed lines are guide lines for induced superconducting transitions at $T_{c1}$ and $T_{c2}$.}
\label{fig5}
\end{figure}

The same $V_{\rm G}$ applied at different temperatures appears to induce different electronic states in the present system. Thus, the $V_{\rm G}$ dependence of $T_{c1}$ and $T_{c2}$ are also sensitive to the temperature $T_{\rm G}$. Here, we select 220, 235, 250(Fig. \ref{fig2}), 280, and 300 K to discuss the role of temperature $T_{\rm G}$ in the induced SC transition. For $T_{\rm G}=$ 220 K, the primary SC transition is broad and crosses over the temperature region between 14 and 11 K as shown in Fig. \ref{fig5}$A$. The SC transition shows no $V_{\rm G}$ dependence, which could be related to the normal-state resistance within the level of k$\Omega$. When increasing the temperature $T_{\rm G}$ to 235 K (Fig. \ref{fig5}$B$), the primary SC transition at $T_{c1}$ becomes steep and the weak transition already appears around 11 K. As shown in Figs. \ref{fig5}$C$ and $D$, the primary SC transition changes from $T_{c1}$ to $T_{c2}$ upon a slight increase of 0.3 V around 3.0 and 2.7 V for $T_{\rm G}=$ 280 and 300 K, respectively, whereas the similar change take places in a wide $V_{\rm G}$ region (3 - 5 V) at 250 K (Fig. \ref{fig2}$B$), suggesting that a higher temperature $T_{\rm G}$ accelerates the evolution of the induced SC phase between $T_{c1}$ and $T_{c2}$. More importantly, it appears that the SC transition at $T_{c1}$ completely disappears when applying $V_{\rm G}\geq$ 3.3 V at 300 K. Upon modifying the temperature $T_{\rm G}$ and voltage $V_{\rm G}$, the systematic behaviors of the induced SC transitions imply the connection and different characteristics between $T_{c1}$ and $T_{c2}$ as the following: (1) similar weak $V_{\rm G}$ dependence, (2) dominant contribution at different $V_{\rm G}$s, and (3) opposite response to reverse $V_{\rm G}$.

\begin{figure}
\centering
\includegraphics[width=1\linewidth]{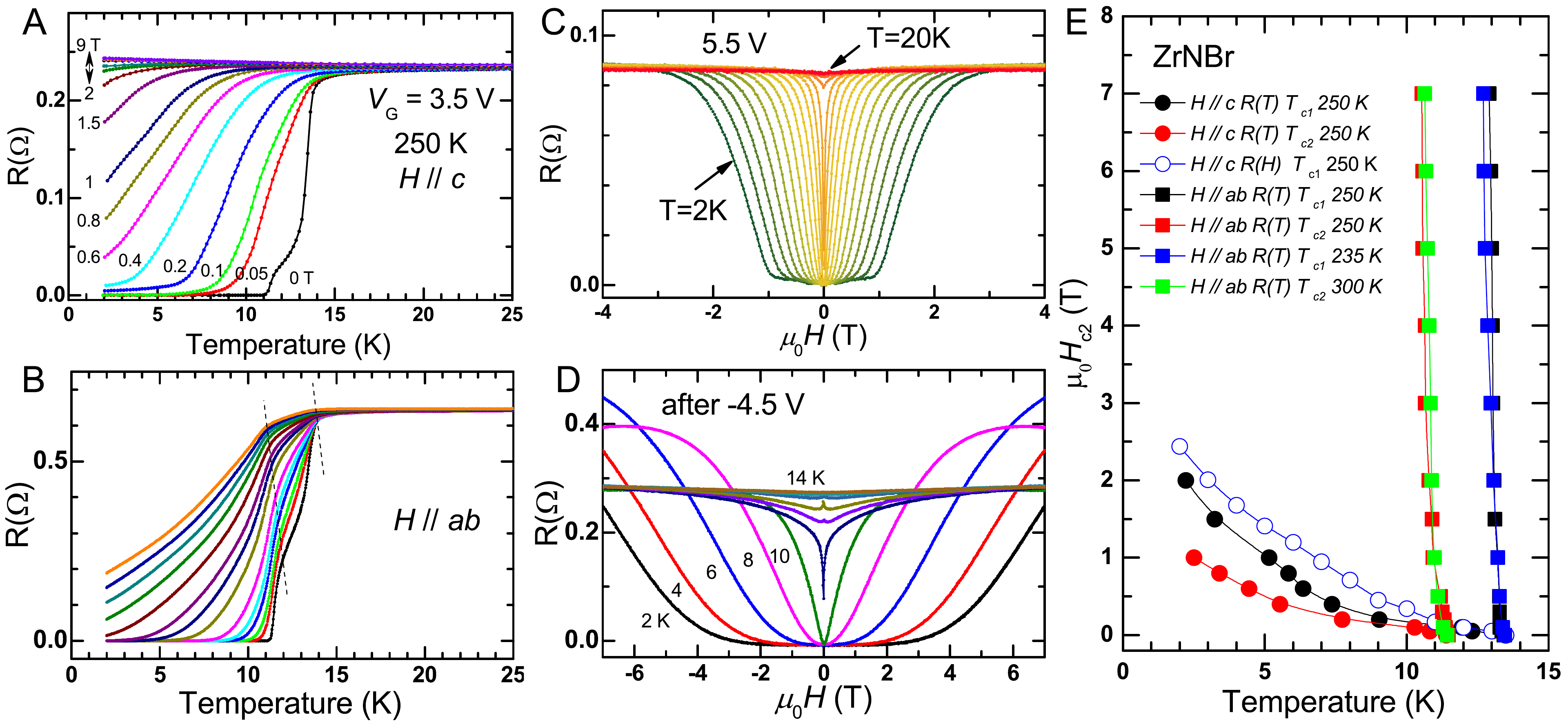}
\caption{Magnetic field dependence of $T_{c1}$ and $T_{c2}$. The temperature dependence of resistance $R(T)$ with $H\parallel c$ ($A$) and $ab$ ($B$) when applying $+3.5$ V at 250 K. For both configurations, magnetic fields are always perpendicular to current flowing in $ab$ plane. For $H\parallel c$, $H$ was increased up to 9 T (0, 0.05, 0.1, 0.2, 0.4, 0.6, 0.8, 1, 1.5, 2, 2.5, 3, 5, 7, and 9 T). For $H\parallel ab$, $H$ was increased up to 7 T (0, 0.05, 0.1, 0.2, 0.3, 0.5, 1, 1.5, 2, 3, 4, 5, 6, and 7 T). The two dashed lines in ($B$) guide the shift of $T_{c1}$ and $T_{c2}$ under fields. Isothermal field dependence of resistance $R(H)$ with $H\parallel c$ ($C$) and $ab$ ($D$) as applying $+5.5$ and $-4.5$ V, respectively. For $V_{\rm G}=$ 5.5 V, $R(H)$ with $H\parallel c$ was measured between 2 and 20 K with an interval of 1 K. $R(H)$ with $H\parallel ab$ was measured at 2, 4, 6, 8, 10, 11.5, 11.75, 12, 12.5, 13, and 14 K. Data of ($A$) and ($B$) were measured on the same single crystal $\#1$ as used in Fig. \ref{fig2}, whereas ($C$) and ($D$) were measured on the single crystal $\#2$. ($E$) The $H_{c2}$-$T$ phase diagram including $T_{c1}$ and $T_{c2}$ values obtained in $R(T)$ and $R(H)$ compared with those of $T_{\rm G}=$ 235 and 300 K.}
\label{fig6}
\end{figure}

Finally, we investigated the magnetic field dependence of $T_{c1}$ and $T_{c2}$ with configurations of $H\parallel c$ and $ab$ as shown in Fig. \ref{fig6}. Here, we select the resistance under fields $R(T,H)$ after applying $V_{\rm G}$s at 250 K as a typical example. $T_{c1}$ and $T_{c2}$ values under magnetic fields are determined using the same method (Fig. \ref{fig2}). As increasing $H$ ($\parallel c$) from 0 to 2.5 T, the SC transition shifts to low temperatures monotonically, and $R(T)$ is almost constant under magnetic fields higher than 3 T as shown in Fig. \ref{fig6}$A$. In contrast, the transitions at $T_{c1}$ and $T_{c2}$ decrease slightly when increasing $H$ $(\parallel ab)$ from 0 to 7 T (Fig. \ref{fig6}$B$). On the other hand, isothermal field dependence of resistance $R(H)$ was also measured to estimate the upper critical field $H_{c2}$ in $c$ and $ab$ as shown in Figs. \ref{fig6}$C$ and $D$, respectively. The method used to determine $H_{c2}$ at certain temperatures is described in Supporting Information section 4. All $T_c$ and $H_{c2}$ values for $T_{\rm G}=$ 250 K are summarized in the $H_{c2}$-$T$ phase diagram as shown in Fig. \ref{fig6}$E$. A huge anisotropy of $H_{c2}$ occurs between $H\parallel c$ and $ab$, implying the 2D SC nature. The SC states mainly induced on the surfaces of the layered ZrNBr crystal are responsible for the strongly enhanced anisotropy of $H_{c2}$ when applying $H$ parallelly along the SC layer. Note that $T_{c1}$ and $T_{c2}$ show comparable field dependence in the configurations of $H\parallel c$ as well as $H\parallel ab$, together with the identical $V_{\rm G}$ dependence (Fig. \ref{fig2}$E$), suggesting the similar SC characteristics. Meanwhile, the correlation between $H_{c2}$ and $T_{c}$ obtained at temperatures $T_{\rm G}=$ 235 and 300 K shows no apparent difference when compared to 250 K. Thus, we can reasonably conclude that the electric field induced superconductivity shows the same characteristic, although the appearance of SC transition at $T_{c2}$ and its disappearance at negative $V_{\rm G}$ and/or high-temperature relaxation can hardly be understood yet. In the present study, we focus on the electric field induced permanent superconductivity and the hidden SC transition at $T_{c2}$, whereas the quantum phase transition recently revealed in 2D ZrNCl and MoS$_2$ is not discussed here\cite{SaitoNC2018}.

The reversible insulating-SC change and permanent superconductivity observed when applying $V_{\rm G}$s at different temperatures suggest that the temperature $T_{\rm G}$ plays a dominant role in electric field induced superconductivity in layered ZrNBr. A small $V_{\rm G}$ (3 V) applied at different $T_{\rm G}$ also induces different electronic states as shown in Supporting Information section 5, implying that the observations are not due to the decomposition of ionic liquid caused by $V_{\rm G}$s that are outside of the EW\cite{SatoEA2004,YuanNL2011}. These phenomena can hardly be fully understood with respect to the general EDLT effect so we attempt to interpret the findings in another scenario. Upon applying $V_{\rm G}$s at 220 K, short-range motion of partial Br$^{1-}$ ions could be induced by the positive $V_{\rm G}$, leading to local electron doping in the system. The application of negative $V_{\rm G}$ and/or long-time relaxation without $V_{\rm G}$ could almost push back the Br ions to their initial positions. This is consistent with the induced SC transition and the insulating state recovered at final 0 V (Fig. \ref{fig4}$D$). Alternatively, we could just use the normal EDLT effect to explain the observations at low $T_{\rm G}$. On the other hand, applying $V_{\rm G}$ at temperatures higher than 250 K could cause  partial Br$^{1-}$ ions to gain enough energy to escape from the crystal surface. The possible formation of Br vacancies can provide extra electrons, which consequently results in a permanent electron doping to the system. The negative $V_{\rm G}$ can only contribute to the improvement of the homogeneity rather than pushing back the Br ions to the material (Fig. \ref{fig4}$F$).

\begin{figure}
\centering
\includegraphics[width=0.9\linewidth]{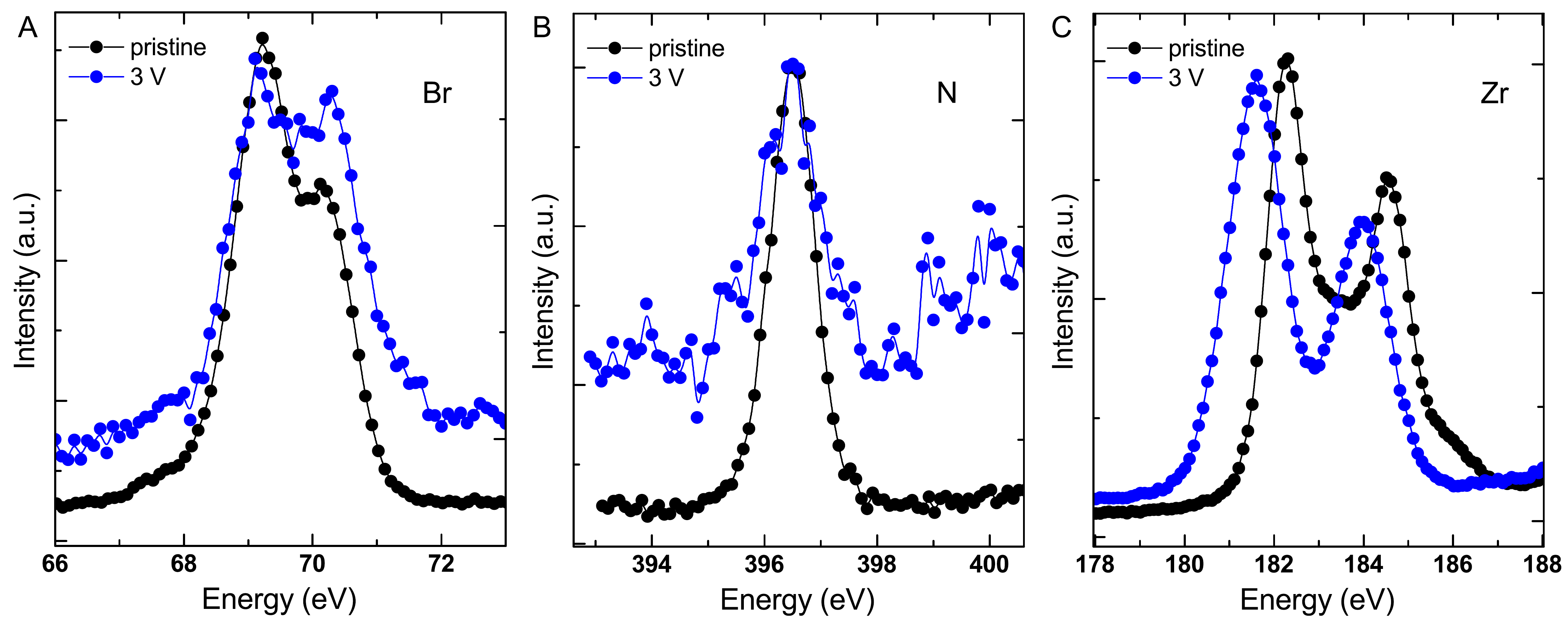}
\caption{XPS spectra of ($A$) Br3d, ($B$) N1s, and ($C$) Zr3d observed in a pristine and gated polycrystalline sample of ZrNBr.}
\label{fig7}
\end{figure}

 The scenario about the Br vacancy formation is supported by the XPS and EDX results from polycrystalline samples. The 3$d$ spectra of Br, N, and Zr are shown in Figs. \ref{fig7}$A$, $B$, and $C$, respectively. Here, our concern is mainly with the difference of binding energy (BE) between before and after the gating process rather than fitting the spectra peaks. To reduce the analysis error, we consider that no shift of the BE of Br occurs after the gating process. Thus, the same correction determined in the comparison of Br BE is applied on those of N and Zr. One can see that the spectra for all three elements become slightly broad after applying 3 V at 300 K, in which the permanent superconductivity has been confirmed. Note that there is a clear shift of near 0.7 eV in the BE of Zr, whereas no shift for the BE of N is observed. More importantly, the BE of Zr shifts to the lower energy region, corresponding to a slight change from oxidation to a sub-oxidation state of Zr after the gating process. Such experimental facts support the permanent electron doping.   
 
\begin{figure}
\centering
\includegraphics[width=0.6\linewidth]{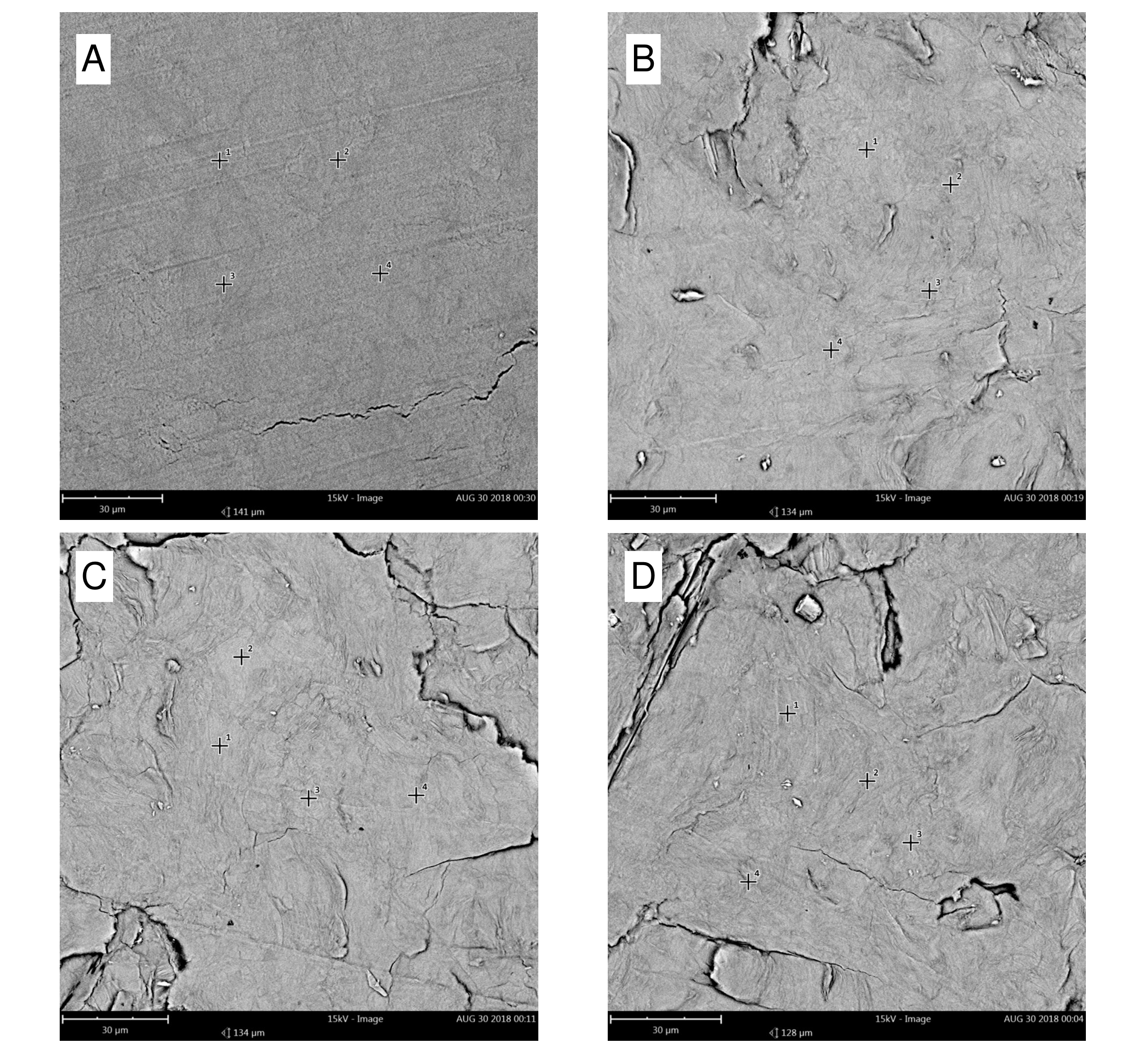}
\caption{The scanning electron microscope image of pristine ($A$) and gated ($B$, $C$, $D$) polycrystalline samples of ZrNBr.}
\label{fig8}
\end{figure}

Meanwhile, the chemical composition of polycrystalline samples was characterized using EDX. The EDX results of one pristine and three gated samples are shown in Figs. \ref{fig8}$A$, $B$, $C$, and $D$, respectively. The average atomic ratio of Zr:Br for the pristine sample is 1 : 0.99$\pm 0.01$, which shifts to 1 : 0.89$\pm 0.01$($B$), 1 : 0.94$\pm 0.01$($C$), and 1 : 0.84$\pm 0.02$($D$) after the gating process. Although this kind of average estimation always has a non-negligible error, one can still consider that the decrease of Br content occurs in the gating process. A similar vacancy forming process was reported in VO$_2$ film, in which suppression of metal-insulator transition was caused by electric field-induced oxygen vacancy formation\cite{JeongScience2013}. Based on the observations in the present study, we reasonably consider that the temperature of 235 K is close to a critical temperature, above which the deintercalation of Br ions could occur when applying proper $V_{\rm G}$s. Thus, the permanent superconductivity observed in the present study could be caused by the partial deintercalation of Br in the gating process.

As shown in Fig. \ref{fig2}$C$, negative $V_{\rm G}$s could result in the increase of $T_c$ and absolute resistance. We consider that the partial deintercalation of Br ions occurs randomly on the crystal surfaces, which induces SC states with poor homogeneity. Upon applying negative $V_{\rm G}$ or long-time relaxation without $V_{\rm G}$ at high temperatures, the local high density of Br vacancies turns into a relatively uniform low density of vacancies, leading to uniform SC states with lower doping levels. Thus, the $T_c$ increases with decreasing carrier density, which is consistent with the observations of bulk superconductors\cite{TakanoPRL2008}. In the meanwhile, the uniform distribution of Br vacancies causes the change from local scattering to relatively uniform scattering, which is responsible for the increase of absolute $R(T)$. We consider that such a process is a dynamic relaxation, especially under negative $V_{\rm G}$s. 

One more result caused by applying the reverse $V_{\rm G}$ is the  disappearance of the SC transition at $T_{c2}$. It is possible that the SC transition at $T_{c2}$ is undetectable in $R(T)$ because of the zero resistance, which is caused by the SC path formed at $T_{c1}$. However, $M(T)$ shows no trace of $T_{c2}$ (Fig. \ref{fig4}$G$), implying that the induced SC transition at $T_{c2}$ really disappears when applying reverse $V_{\rm G}$ and/or relaxation at high temperatures without $V_{\rm G}$. Although the SC phase with $T_{c2}$ is very similar to a conventional EDLT induced high doping level, we consider that the SC state with $T_{c2}$ could be a kind of metastable state with energy higher than the bulk stable state. When applying positive $V_{\rm G}$, local carriers present at both the stable and the metastable states, inducing SC states with $T_{c1}$ and $T_{c2}$, respectively. It appears that a higher $V_{\rm G}$ could result in a dominant contribution from the metastable states to the transport channel. When applying a reverse $V_{\rm G}$, the local carriers at the metastable energy state could be transferred to the stable states, which results in the stable states corresponding to the primary SC phase with $T_{c1}$. One can see some similarity between the electric field induced metastable SC states discussed here and the gate-induced incommensurate/nearly commensurate current density wave phase transition (ICCDW/NCCDW) observed in 1T-TaS$_2$\cite{YuNN2015}. 
However, the origin of such a metastable state is yet unclear.

In conclusion, we find that the processing temperature $T_{\rm G}$ as well as the gate voltage $V_{\rm G}$ plays a dominant role in electric field induced superconductivity in layered ZrNBr. $T_{\rm G}=$ 235 K appears to be a boundary between the conventional EDLT effect and electric field induced electrochemical reaction using DEME-TFSI. We consider that such a temperature effect is not limited to a certain type ionic liquid. The permanent superconductivity is caused by the partial Br vacancy, which is supported by the irreversible insulating-SC change when applying reverse $V_{\rm G}$, the shift of BE of Zr, and the decrease of Br content on the crystal surfaces. Although the mechanism is not clear yet, a metastable SC phase with $T_c=$ 11 K has been revealed to coexist with the primary SC transition at 14 K. More exotic phenomena could be expected upon modifying the temperature $T_{\rm G}$ and voltage $V_{\rm G}$ using an EDLT device with liquid dielectric.    

\section*{methods}
Pristine ZrNBr single crystals were grown using a well-established chemical transport method\cite{Yamanaka1998}. Typical crystal size is 300 $\times$ 200 $\times$ 10 $\mu m^3$. Layered crystals were fixed on a SiO$_2$ surface grown on a Si substrate. A schematic configuration of the device is described in detail in a previous study\cite{ZhangCPL2018} and in Supporting Information section 1. Compared to conventional field effect transistors, no drain voltage was applied when applying $V_{\rm G}$s between gate and source, and no terminal was connected to ground. Ionic liquid of Diethylmethyl (2-methoxyethyl) ammonium bis (trifluoromethylsulfonyl) imide (DEME-TFSI) and 1-Ethyl-3-methylimidazolium tetrafluoroborate (EMIM-BF$_4$) were chosen as the gate dielectric. Proper $V_{\rm G}$s were applied at several temperatures between 220 and 300 K. All processes of applying $V_{\rm G}$ were performed in a vacuum to avoid the influence of oxygen and water in the air. The applied $V_{\rm G}$ was maintained during the cooling down of the system and released after the appearance of $\pm$ 0.1 nA of leakage current. The temperature dependence of electrical resistance $R(T)$ discussed in the present study was measured in a warming up process. At each temperature selected for applying $V_{\rm G}$s, systematic measurements of $R(T)$ were performed on a new crystal. After the measurement of $R(T)$ with the highest $V_{\rm G}$ around the EW limit at each process temperature $T_{\rm G}$, the system was kept at the same temperature without $V_{\rm G}$ for 1 - 2 hours. The $R(T)$ measured after the relaxation process is named final 0 V, which was followed by the $R(T)$ with a reverse $V_{\rm G}$. After completing all resistance measurements, the temperature dependence of the magnetization $M(T)$ on the same single crystal was measured using a SQUID magnetometer (Quantum Design SQUID VSM). The XPS measurements were performed on a Thermo Scientific ESCALAB 250X spectrometer fitted with a monochromatic Al $K_{\alpha}$ X-ray source. The composition of polycrystalline samples were estimated using Oxford X-Max energy dispersive X-ray spectroscopy (EDX). 

\section*{Data availability}
All data are available upon request from the corresponding authors.
\section*{Acknowledgments}
This work was supported by the National Natural Science Foundation of China (Grant No. 11704403), the National Key Research Program of China (Grant No. 2016YFA0401000 and 2016YFA0300604), the Strategic Priority Research Program (B) of Chinese Academy of Sciences (Grant No. XDB07020100) and the National Basic Research Program of China 973 program (Grant No. 2015CB921303).
S.Z conceived the project and designed the experiments. X.W. and H.F. fabricated the polycrystal samples and single crystals of ZrNBr. S.Z. designed and fabricated the EDLT devices. X.W., H.F., S.Z. and M.G. carried out all of the measurements. S.Z., X.W. and H.F. analysed the data. S.Z wrote the paper, and all authors commented on it.

The authors declare no competing interests.



\end{document}